\documentclass[reprint,showpacs, aps]{revtex4}
\usepackage[utf8]{inputenc}
\usepackage{natbib}
\usepackage{graphicx}
\usepackage{longtable}
\usepackage{subfigure}
\usepackage[tbtags]{amsmath}
\usepackage{booktabs}
\usepackage{makecell}
\usepackage{diagbox}
\usepackage{lastpage}
\usepackage{amssymb,bm,mathrsfs,bbm,amscd}

\usepackage{dcolumn,booktabs}
\newcolumntype{d}[1]{D{.}{.}{#1}}
\newcommand\mc[1]{\multicolumn{1}{c}{#1}} 

\begin{document}

\title{Benchmark study of Nagaoka ferromagnetism by spin-adapted full configuration interaction quantum Monte Carlo}

\author{Sujun Yun}
\email{s.yun@fkf.mpg.de}
\affiliation{Max Planck Institute for Solid State Research, Heisenbergstr. 1, 70569 Stuttgart, Germany}
\affiliation{School of electronic engineering, Nanjing XiaoZhuang University, Hongjing Road, Nanjing 211171, China}

\author{Werner Dobrautz}
\email{w.dobrautz@fkf.mpg.de}
\affiliation{Max Planck Institute for Solid State Research, Heisenbergstr. 1, 70569 Stuttgart, Germany}%

\author{Hongjun Luo}
\email{h.luo@fkf.mpg.de}
\affiliation{Max Planck Institute for Solid State Research, Heisenbergstr. 1, 70569 Stuttgart, Germany}

\author{Ali Alavi}
\email{a.alavi@fkf.mpg.de}
\affiliation{Max Planck Institute for Solid State Research, Heisenbergstr. 1, 70569 Stuttgart, Germany}%
\affiliation{Yusuf Hamied Department of Chemistry, University of Cambridge, Lensfield Road, Cambridge CB2 1EW, United Kingdom}%

\begin{abstract}

We investigate Nagaoka ferromagnetism in the two-dimensional Hubbard model with one hole using 
the spin-adapted ($SU(2)$ conserving) full configuration interaction quantum Monte Carlo method.
This methodology gives us access to the ground state energies of all possible spin states $S$ of finite Hubbard lattices, here obtained for lattices up to 24 sites, for various interaction strengths ($U$). 
The critical interaction strength, $U_c$, at which the Nagaoka transition occurs is determined for each lattice and is found to be proportional to the lattice size for the larger lattices. 
Below $U_c$ the overall ground states 
are found to favour the minimal total spin ($S=\frac 1 2$), and no intermediate spin state is found to be the overall ground state on lattices larger than 16 sites. However, at $U_c$, the energies of all the spin states are found to be nearly degenerate, 
implying that large fluctuations in total spin can be expected in the vicinity of the Nagaoka transition.
\end{abstract}


\maketitle

\section{Introduction}
The two-dimensional Hubbard model is an important theoretical model in condensed matter physics, and exact results are helpful for understanding a plethora of phenomena in strongly correlated systems, including 
pairing mechanisms in unconventional superconductors
\citep{scalapino2012}, the Mott metal-insulator transition \,\cite{imada1998} and magnetism. The magnetic properties of the ground state wavefunctions, off half-filling, are still an open problem. The first known example of saturated itinerant electron ferromagnetism is due to Thouless\cite{Thouless1965} for some special bipartite lattices, and was later generalized and applied to non-bipartite lattices by Nagaoka \cite{Nagaoka1966}, Lieb\cite{lib1989} and Tasaki\cite{Tasaki1989,Tasaki1998}, for systems containing exactly one hole with an infinite Hubbard repulsion. Today, this phenomenon is known as Nagaoka ferromagnetism.

However, a detailed physical picture of the phase transition point is still not clear; for example it is not known if there are states with intermediate spin which are particularly stable in the vicinity of the phase transition from anti-ferromagnetism to ferromagnetism. 
Furthermore, the existence of Nagaoka ferromagnetism in the case of more than one hole is controversial.
Nagaoka ferromagnetism plays an important role in the study of the magnetic properties of the Hubbard model, because it states that there is a ferromagnetic ground state in the vicinity of half filling, where an anti-ferromagnetic spin order is assumed to be present. It is also a rigorous result reporting ferromagnetism in Hubbard model. 
Nagaoka proved that the ground state of certain Hubbard models have saturated ferromagnetism, if there is one hole and $U=\infty$. 
This theorem, however, does not offer a picture on how the system changes from an anti-ferromagnetic state at small $U$ to a saturated ferromagnetic state at $U=\infty$. 
To the best of the authors' knowledge, there exists no work on locating the Nagaoka  critical strength $U_c$ on finite lattices, and studying the physical properties on such lattices.
In this work we investigate the energetics of different spin states in the Nagaoka problem on finite lattices and obtain insight into the spin spectrum as $U_c$ is approached. 
Besides, this problem also offers an extreme example of a strongly-correlated itinerant system, which can be used to test and improve newly developed methods for strong correlations.

In order to get the most reliable benchmark results on strongly correlated systems, one usually needs to use various kinds of highly accurate methods. 
The Lanczos-based exact diagonalization (ED) method is computationally very expensive, especially the demand for memory is extremely high. For the Hubbard model, it is generally prohibitive to use the ED method on lattices larger than 18 sites \cite{Lin1991,becca2000}. 
The recently developed full configuration interaction quantum Monte Carlo (FCIQMC) method is capable of generating highly accurate benchmark results with a lower memory requirement than ED, even in its original formulation \cite{Booth2009}, where no initiator approximation \cite{cleland2010} is applied. 
The FCIQMC method and its initiator approximation have been tested on various molecular systems \cite{blunt2015,thomas2015,khaldoon2019} as well as on some lattice \cite{Schwarz2015, Werner2019} and solid \cite{booth2013,luo2018} systems.
The ground state wave function of such systems usually contains a relatively small number (typically not more than a few thousand) of important reference determinants, which form the dominant part of the wave function.  This part of the wave function is normally very stable in the course of the Monte Carlo simulation, and can be used as a trial wave function to project out the ground state energy. This, however, is not the case in the Nagaoka ferromagnetic states, where all determinants containing no double occupation are equally important and it is very difficult to get a stable projection energy, especially when $U$ is close to $U_c$.
Therefore, Nagaoka ferromagnetism is a challenging problem for this methodology.

In FCIQMC, the basis is usually formulated in the Hilbert space of Slater determinants, 
where each individual determinant is an eigenstate of $\hat S_z$, but generally not of the square of the total spin operator, $\hat{{\bf S}}^2$. To study magnetism, especially in systems with small spin-gaps, it is useful to impose the $SU(2)$ symmetry arising from the commutator $ [\hat{H},\hat{{\bf S}}^2]=0$, i.e effectively  having a basis with a specific total spin. This enables one to target specific spin states which are not necessarily the ground state (for example intermediate spin states), and in projective methods also leads to faster convergence, as unwanted spin components are rigorously absent from the Hilbert space. It also helps to reduce the size of the Hilbert space.  
As a price for these advantages, one has to construct such a spin-adapted basis in a sophisticated way.
In this paper, $SU(2)$ symmetry is imposed via the Graphical Unitary Group Approach (GUGA) \cite{Paldus1976,Shavitt1977,Shavitt1978} which dynamically constrains the total spin $S$ of a multiconfigurational and highly open-shell wavefunction in an efficacious manner. 
Recently, a spin-adapted version of the FCIQMC algorithm based on GUGA has been developed in our group \cite{Wernerandsimon2019,Giovanni2020}, with applications so far only to ab initio systems. In this paper we report an application of the GUGA-FCIQMC method to the Hubbard model in the 
large $U$ regime, where spin gaps are very small. To our knowledge, this is the first time that an exact spin-adapted methodology has been applied to the Nagaoka problem. 

The rest of this paper is organized as follows.
In Sec. \ref{method}, the GUGA-FCIQMC method is briefly introduced.  In Sec. \ref{results}, we present the computation results on various lattices with this method.  In Sec. \ref{conclusion}, we make some conclusions of this work.

\section{ GUGA-FCIQMC method applied to the real-space Hubbard model }
\label{method}

The Hamiltonian of the Hubbard model in real space takes the form
\begin{equation}
\hat{H} = - t \sum_{\langle i j \rangle \sigma}a^{\dagger}_{i, \sigma} a_{j, \sigma} + U \sum_{i} n_{i\sigma} n_{i\overline{\sigma}} \label{oriHamil}
\end{equation}
where $a^{\dagger}_{i \sigma}$ ($a_{i \sigma}$) creates (annihilates) an electron with spin
$\sigma$ on site $i$ and $ n_{i\sigma}$ is the particle number operator.
$U$ refers to the Coulomb interaction strength. We consider only nearest neighbour hopping terms, where $t$ is positive and which we use as the unit of the energy. 

This model represents an itinerant strongly correlated systems, especially in the large $U$ regime off half-filling. To get reliable ED-quality results on such strongly correlated systems, the recently developed FCIQMC method is used to obtain the ground state wave function $\Psi_0$  by Monte Carlo simulation of the imaginary-time evolution of wave functions \begin{equation}
\Psi (\tau) = e^{-\tau (\hat{H}-E_0 )}\Psi(0) 
\label{eq_Psi_t}
\end{equation}
which leads to $\Psi_0$ in the long time limit $\Psi(\tau \rightarrow \infty)$. If the initial wavefunction $\Psi(0)$ has a definite spin $S$ (which may be different to the spin of the true ground state), this procedure in principle leads to the lowest energy state of that spin. This however requires that the imaginary-time propagation of the wavefunction rigorously preserves the spin from one iteration to the next, otherwise any noise leads to the collapse of the wavefunction onto true ground state with a possibly different spin, and the desired spin state remains inaccessible. The exact preservation of spin is a major challenge for stochastic projection techniques working in Slater determinant spaces, since the Slater determinants are generally not individually spin eigenfunctions. For this reason, special algorithms such as the GUGA-FCIQMC algorithm need to be devised, in which the spin is rigorously preserved even in a stochastic simulation.  

The wave function is expressed in terms of a complete basis of spin eigenfunctions $\{|\mu\rangle\}$
\begin{equation}
\Psi(\tau) = \sum_\mu c_\mu(\tau) |\mu\rangle
\end{equation}
The coefficients $c_\mu$ are determined via a population dynamics of signed walkers, $s_\nu \delta(\nu-\mu), s_\nu=\pm 1$, such that $N_\mu=\sum_\nu s_\nu \delta(\mu-\nu)$ represents the population of walkers on $|\mu\rangle$. This population dynamics follows the master equation: 
\begin{equation}
-\frac {d N_\mu} {d \tau} = (H_{\mu\mu}-E)N_\mu+\sum_{\nu\neq \mu} H_{\mu \nu} N_\nu
\label{master}
\end{equation}
Here $H_{\mu\nu}=\langle \mu|\hat{H}|\nu \rangle$ is a matrix element of $\hat{H}$ in the given basis. The efficient on-the-fly evaluation of such matrix elements is key in any iterative method, and in a spin-adapted basis, this forms the main technical problem to be overcome. The parameter $E$, called the shift parameter, plays an important role to control the population growth and converges to the ground state energy $E_0$ in the long time limit. 

In a spin-adapted method, the basis functions $\{|\mu\rangle\}$ are chosen to be eigenstates of $\hat{{\bf S}}^2$ and $\hat{S}_z$. Expanded in a Slater determinant basis, such spin eigenfunctions generally entail a combinatorially large number of Slater determinants, dependent on the total spin and the number of singly-occupied orbitals in the constituent SDs. In the large $U$ Hubbard model, the latter is essentially the number of electrons (there are very few doubly occupied sites) and therefore spin-adapted bases are extremely multi-determinantal in Nagaoka-type problems. For this reason, one must seek methods in which matrix-element calculation can be performed directly, rather than via expansions in Slater determinants. The GUGA approach is one such approach, that uses the algebra of the Unitary group to perform efficient matrix-element calculation and below we briefly describe the basis of this method. 

The Hubbard Hamiltonian can be reformulated in terms of spin-free excitation operators as
\begin{equation}
\hat{H} = -t\sum_{\langle i j \rangle} \hat{E}_{ij} + \frac{U}{2} \sum_{i}\hat{e}_{ii,ii} \label{Hamil}
\end{equation}
where the sums $i,j$ run over the $N_s$ lattice sites, and 
\begin{equation}
 \hat{E}_{ij}=\sum_{\sigma=\uparrow,\downarrow}a^{\dagger}_{i, \sigma} a_{j, \sigma}  \nonumber
\end{equation} 
and
 \begin{equation}
 \hat{e}_{ii,ii}=\sum_{\sigma}a^{\dagger}_{i, \sigma} a^{\dagger}_{i, \overline{\sigma}} a_{i,\overline{\sigma}} a_{i, \sigma}
\end{equation}
are singlet one-body and two-body excitation operators. Since these excitation operators commute with the total spin operator $\hat{{\bf S}}^2$ and the $z$-component $\hat{S}_z$, they preserve the $S$ and $S_z$ values upon acting on a spin eigenstate $|S,S_z\rangle$.
Because the spin-free excitation operators in Eq.(\ref{Hamil}) obeys the same commutation relations as the generators of the Unitary Group $U(n)$ ($n=N_s$ being the number of spatial orbitals),
$U(n)$ can be used to construct a spin-adapted basis, also known as configuration state functions (CSFs), via the Gel'fand-Tsetlin (GT) representation of $U(n)$ \cite{Paldus1976}. This formalism results in the {\em Graphical Unitary Group Approach} method, and can be applied to the Hubbard model in the form given in Eq. (\ref{Hamil}). Paldus \cite{Paldus2020} has given a detailed derivation of the matrix elements of the Unitary group generators in the GT basis, and for the implementation of the GUGA formalism within the stochastic FCIQMC framework the reader is referred to Ref.\,\cite{Wernerandsimon2019}. For the purposes of this study, it should be noted that, because of the simple form of the Hubbard Hamiltonian, only a small subset of the possible GUGA matrix elements are necessary to be calculated here, and fortunately the required ones are relatively simple compared to the general forms which are necessary for ab initio Hamiltonians. Thus, for the off-diagonal matrix elements, $\langle\mu^\prime|\hat{H}|\mu \rangle$, only the one-body terms contribute and necessary GUGA matrix elements are of the form 
$\langle\mu^\prime|\hat{E}_{ij}|\mu \rangle$, which is given in Appendix A of the aforementioned paper. The diagonal matrix elements $\langle\mu|\hat{H}|\mu \rangle$ require only the GUGA matrix elements of the form  $\langle \mu^\prime|\hat{e}_{ii,ii}|\mu \rangle$, whose expression is given in equation B4 of Appendix B of Ref.\,\cite{Wernerandsimon2019}, and can be calculated in $O(N_s)$ effort. 

In this spin-adapted formalism, the dimension $f$ of the Hilbert space of a system with $N_s$ sites, $N_e$ electrons and spin $S$ is given by the Weyl-Paldus formula \cite{Paldus1976}:
\begin{equation}
    f(N_s,N_e,S) = \frac{2S+1}{N_s+1} \binom{N_s+1}{N_e/2-S} \binom{N_s+1}{N_s-N_e/2-S}
\end{equation}
In this study, we use the GUGA-FCIQMC method up to $N_s=24$. The corresponding largest Hilbert space results for $N_e=23, S=3/2$, i.e. $f\sim 2.3\times 10^{12}$. This would be the dimension that an exact diagonalisation method would need to allocate to store the ground-state eigenvector. Such calculations would only be feasible with specialised code on supercomputers with large amounts of memory.   

The Hilbert space associated with the no-double occupancy sector $f_{ND}$ is much smaller, and this is where the majority of the ground-state eigenvector in the large $U$ limit resides. For the one-hole Nagaoka problem, $f_{ND}$ is given by the Sherman van-Vleck formula   \cite{vanVleck1935}
multiplied by the number of sites:
\begin{equation}
  f_{ND}(N_s,N_e,S) = N_s \times \left( \binom{N_e}{N_e/2-S}-\binom{N_e}{N_e/2-S-1}\right)
\end{equation}
For the 24-site lattice with 23 electrons and $S=\frac{3}{2}$, $f_{ND}\sim 8\times 10^6$. In the full GUGA-FCIQMC method reported in this study (i.e. without the initiator approximation), the number of walkers required to resolve the sign-problem for the Nagaoka-type problems is found to be roughly $5-10$ times $f_{ND}$, i.e. on the order of $10^8$ walkers, which is still considerably less than the $10^{12}$ Hilbert space an exact deterministic spin-adapted calculation would need in order to solve this problem. It is this saving that makes these essentially exact calculations possible on a medium-size machine (several tens of  processors). 

\section{results }
\label{results}

In this work we investigate the 2D Hubbard model on square lattices with periodic boundary conditions, where the existence of the Nagaoka ferromagnetism has been proven for the case of $t>0$ and $U=\infty$. 
Calculations are mainly performed on lattices of three different sizes. Beside a simple 16-site ($4\times 4$) square lattice, we also take two other square lattices under the tilted periodic boundary conditions, which are the 18-site square lattice with lattice vectors $(3,3), (-3,3)$ and the 24-site square lattice with lattice vectors $(5,1), (1,5)$. These two tilted square lattices have optimum shape for finite clusters, and help to decrease finite size effects\cite{betts1999}.  
\begin{table*}[htbp] 
  \centering \caption{\label{tab:1} Comparison between  the spin dependent ground state energies calculated by GUGA-FCIQMC and those by the exact diagonalization  on the 16-site square lattice Hubbard model.  The number of electrons is 15 (one hole) and the Hubbard repulsion parameter is taken as $U=60$.}
 \begin{tabular}  {llllllllcccc}\hline  \hline
  \diagbox[width=4cm] {Method}{$S_{total}$} & \mc{$\frac {1} {2}$}   & \mc{$\frac {3} {2}$}  & \mc{$\frac {5} {2}$}  & \mc{$\frac {7} {2}$}  & \mc{$\frac {9} {2}$} & \mc{$\frac{11}{2}$} & \mc{$\frac{13}{2}$}  \\  \hline
  GUGA-FCIQMC  &$-4.0724(1)$  &  $-4.0690(1)$ & $-4.0753(1)$ &  $-4.0662(1)$ &  $-4.0138(1)$  &  $-4.0195(1)$  & $-3.9949(1)$ \\
  Exact diagonalization &$-4.07242$ &  $-4.06923$    &$-4.07535$  &$-4.06603$  &$-4.01379$   &$-4.019409$ & $-3.994902$ \\
     \hline \hline
  \end{tabular}
  \begin{flushleft}
  \end{flushleft}
 \end{table*}

To benchmark the performance of the GUGA-FCIQMC method on Hubbard model, we first apply this method to a $4\times4$ lattice with one-hole and compare the results directly with those of Lanzcos-based ED method in Slater determinant space in different $S_z$ sectors starting from $S_z=\mbox{maximal}$ and reducing $S_z$ successively to zero, and calculating multiple roots of the many-body Hamiltonian. The spin of each root can retrospectively assigned. 
The results are presented in Table~\ref{tab:1}, where $U=60$ is taken for the test. In the table the results for the $S_{max}$ ($=\frac{15}{2}$), which is the Nagaoka ferromagnetic state, is not shown since the ground state energy of this state is identically equal to $-4$, and can expressed by a small multi-configurational wavefunction. For all the different spin states, the results of the two methods agree extremely well, to within the stochastic error of $\sim 10^{-4}$, and confirm the correct implementation of the GUGA-FCIQMC methodology. 

 \begin{figure}[htbp]
\centering
\includegraphics[width=5cm]{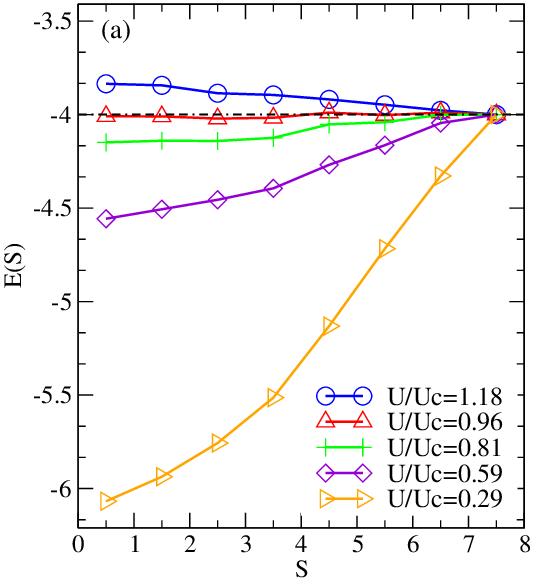}
\vspace{0.5cm}
\hspace{0.5cm}
\includegraphics[width=5cm]{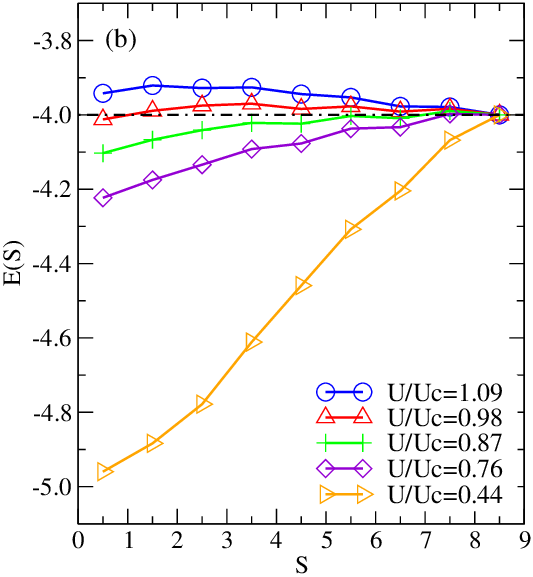}
\vspace{0.5cm}
\hspace{0.5cm}
\includegraphics[width=5cm]{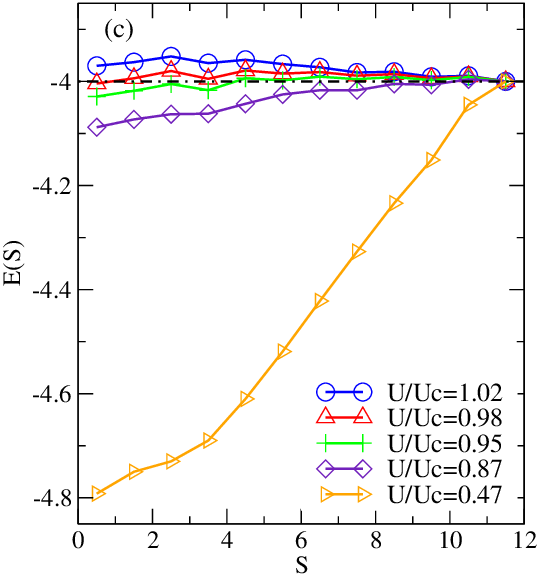}
\caption{Ground state energy $E$ (in units of $|t|$) versus the total spin $S$ on (a) 16, (b) 18  and (c) 24 site lattices with one hole. In the plots, the $U$'s are referred to the $U_c$, which are taken as 68, 92 and 127 for the three lattices. This means that when $U/U_c=1$, the energy of the $S=\frac 1 2$ state is equal to within error bars to that of the maximal spin. 
The lines are a guide to the eye.} 
\label{fig1}
\end{figure}

In Figure~\ref{fig1},  the results of the ground state energies $E(S)$ of systems with one hole with spin $S$ are presented, for the 16, 18 and 24 site square lattices, for different values of $U$. The value of $U=U_c$ at which the maximal spin state becomes the ground state locates the Nagaoka transition, and is numerically found to be strongly dependent on the lattice size, namely $U_c=68, 92$ and $127$ respectively, for the 16, 18 and 24 site square lattices. In order to compare the behaviour of the $E(S)$ for the different lattices at different $U$, the displayed values of $U$ are normalised to these $U_c$'s. The behaviour of the functions $E(S)$ is quite similar for the three lattices. At small interaction strength $U$, the ground state energy with $S=S_{min}=\frac 1 2$ takes the lowest value among all different spins, with a monotonic increase in energy with $S$. With increasing $U$, the shape of this curve steadily flattens, with 
energy $E(\frac 1 2)$ rising up and finally exceeding the ground state energy at the maximum spin $E(S_{max})=-4$. It is observed that in the vicinity of $U/U_c=1$, the function $E(S)$ is almost flat for all $S$, implying near degeneracy of all spin states.

It is interesting to ask how this phase transition takes place: whether it is a sudden jump from a state with $S=\frac 1 2$ to the state with $S=S_{max}$, or if it is more gradual, i.e. if there exists a regime of $U$ where the lowest energy state takes an intermediate value of spin $\frac 1 2 <S<S_{max}$. On the 16-site lattice, we find that at $U/U_{c}=0.81$ and $U/U_{c}=0.96$ the states with the lowest energies have an intermediate spin $S=\frac 5 2$. On the two larger lattices, however, we find at all different interaction strengths, the states with the lowest energies take either the minimum spin ($S=\frac 1 2$) or the maximum spin ($S=S_{max}$). 
We regard the existence of an intermediate spin ground-state on the 16-site square lattice as an artifact of the small lattice size.

\begin{figure}[htbp]
\centering
\includegraphics[width=4.8cm]{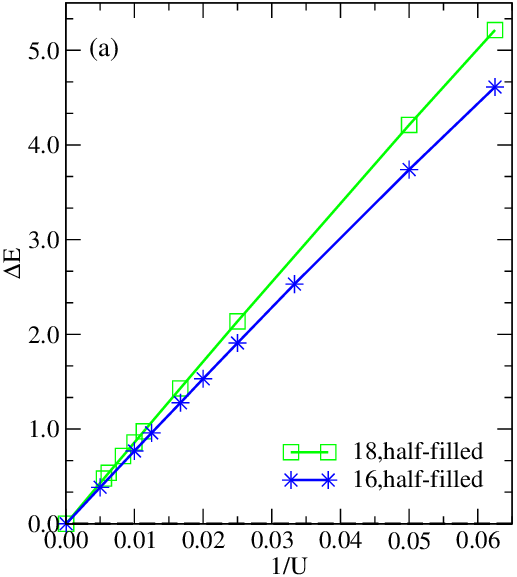}
\vspace{0.5cm}
\hspace{0.5cm}
\includegraphics[width=4.8cm]{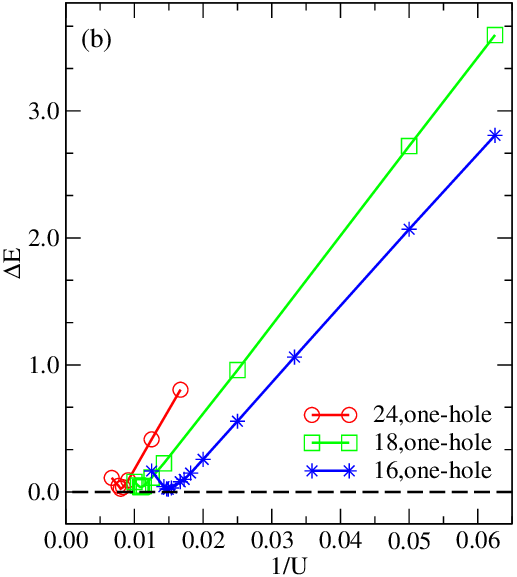}
\vspace{0.5cm}
\hspace{0.5cm}
\includegraphics[width=4.8cm]{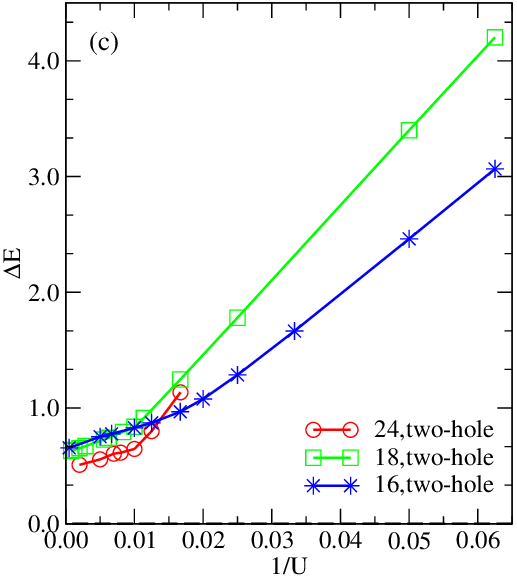}
\begin{flushleft}
\caption{\label{width} The width of the spin spectrum $\Delta E$ versus $1/U$ in case of (a) half-filled, (b) one-hole and (c) two-holes, on different lattices.
The lines are a guide the eye and the statistical errors of $\Delta E$ can not be seen on this scale.}
\end{flushleft}
\end{figure}

To get some more insight into these results, we also measure the {\em width} of spin spectrum $\Delta E=E_{max}-E_{min}$ for every given strength $U$, where $E_{max}= \max(E(S)), E_{min}=\min (E(S))$ are the maximal and minimal values of the energy over all spin states for a given $U$.
In Figure~\ref{width}, the results of $\Delta E$ are plotted as functions of $1/U$, for the half-filled, one-hole and two-hole systems respectively in the various lattices.
For the half-filled system(see Fig.2(a)), $\Delta E$ is simply proportional to $U^{-1}$, and it converges to 0 in the
limit $U^{-1}\rightarrow 0$. This result is expected, namely only at $U=\infty$ is the system fully spin degenerate. 


The situation for the one-hole system, see Fig.~\ref{width}b, is quite different. As a function of decreasing $U^{-1}$ (increasing $U$), the spin spectrum gets linearly reduced, achieving a small value at $U_c$, and then in a cusp-like manner increasing again. 
The quality of the linear fit is striking, and this makes it possible to locate the phase transition point $U_c$ with a few calculations in a $U$ regime away from the difficult $U_c$ point.  At $U_c$ the width $\Delta E$ gets a minimum value, which is measured as $<0.03$, meaning that at the transition point the ground state energies of all the spin states are nearly degenerate to within this energy window. This implies that very large spin fluctuations can be expected in the vicinity of the Nagaoka transition. A further consequence of this massive near-degeneracy should be non-trivial behaviour of the entropy and heat capacity through this transition. This however would be best studied using a finite-temperature method \cite{Blunt2014,Blunt2015b} rather than a ground-state technique. 

\vspace{1cm}
\hspace{1cm}
\begin{figure}[htbp]
\centering
\includegraphics[width=6.5cm]{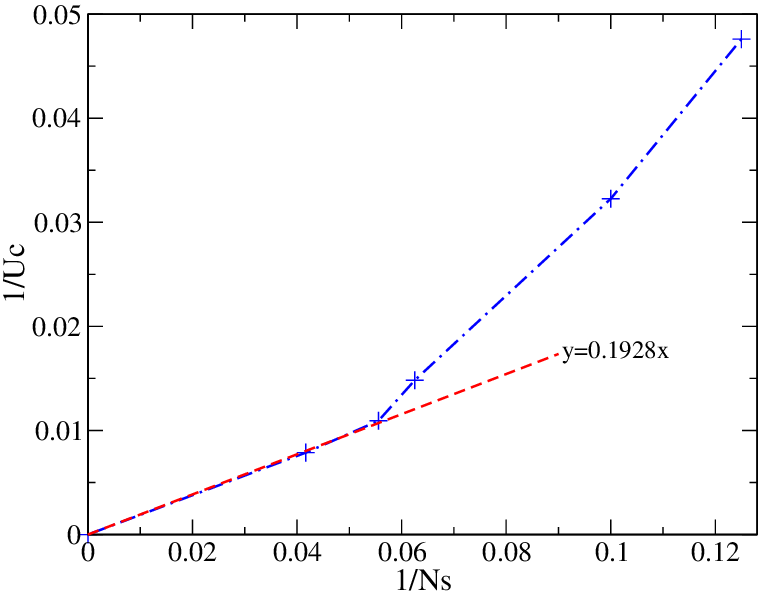}
\begin{flushleft}
\caption{\label{Uc} The inverse of the critical interaction strength, $1/U_c$, versus the inverse 
of the lattice size, $1/N_s$. The $(0,0)$ point is assumed to be the Nagaoka result at the $N_s=\infty$  and $U=\infty$ limit. This point is extremely well-extrapolated from the 18 and 24-site results (red line). This demonstrates that $U_c$ is expected to scale linearly with $N_s$ for square lattices larger than 18 sites.}
\end{flushleft}
\end{figure}

To study the dependence of $U_c$ on the lattice size $N_s$, we plot the inverse of the transition strength ($1/U_c$) as a function of the inverse of the lattice size ($1/N_s$) in Figure~\ref{Uc}. On small lattices, such as the 8, 10 and 16 site square lattices, there is no clear relationship between $1/U_c$ and $1/N_s$. This may be understood as the consequence of strong finite size effects. On the other hand, we find that the linear extrapolation of the two results on 18 and 24 site square lattices point to $(0,0)$, where the result for the large $N_s$ limit should be located (i.e., for $1/N_s\rightarrow 0$, $1/U_c\rightarrow 0$). This already gives us a simple linear relation between $1/U_c$ and $1/N_s$ in the asymptotic large lattice regime, numerically found to be  $1/U_{c}=0.1928/N_{s}$, which means that the $U_c$ grows linearly with system size, achieving the value of infinity in the limit of infinite $N_s$, consistent with the Nagaoka theorem. 

The situation with two holes, see Fig.~\ref{width}c, is also interesting. There, the spin spectrum retains a finite width even at very large values of $U$, with the $S=0$ state remaining the ground state with  a noticeable gap in these finite systems to higher spin states. This implies that the type of spin near-degeneracy observed in the one-hole system does not occur in the two-hole case. However, we also observe a clear change in slope in the width of the spin spectrum as the system passes through $U_c$. Preliminary analysis of this behaviour indicates the onset of ferromagnetic domains in the two-hole system, which are anti-ferromagnetically aligned with respect to each other, leading to an overall $S=0$ ground state. This indicates that Nagaoka physics also remains in multiple-hole systems, but is much more complicated due to effective interactions between different ferromagnetic domains. This is a topic we hope to address in a future publication. 
\section{Conclusion}
\label{conclusion}
The spin-adapted full configuration interaction quantum Monte Carlo via the graphical unitary group approach (FCIQMC-GUGA) is used to study Nagaoka ferromagnetism of 2D Hubbard model with one hole on finite lattices. The largest lattice is up to 24 sites where the finite size effect in the large U regime is very weak. Based on the results, we find that below the phase transition strength $U_c$ the ground states always prefer the minimum total spin $S=\frac 1 2$, and there is no partial spin polarization on square lattices larger than 16 sites. At the phase transition strength $U_c$ the ground-states becomes nearly degenerate among all different spins. The results also show that the phase transition strength ($U_c$) is proportional to the lattice size ($N_s$).
The present methodology can be extended to the calculation of reduced density matrices, giving insight into the spatial and spin correlations in the observed wavefunctions, and will be reported in a future publication. 


\begin{center}
{\large Acknowledgments }
\end{center}
Olle Gunnarsson is thanked for providing the exact diagonalisation results quoted in this paper. 
Sujun Yun  would like to extend her sincere gratitude
to Vamshi Katukuri and Kai Guther of the Max Planck Institute for Solid State Research, 
Youjin Deng in the University of Science and Technology of China and Qianghua Wang in Nanjing University for participating in discussion and providing support. 
The authors gratefully acknowledge funding from the MPG. 
Sujun Yun is supported by the national natural science foundation of China (Grant No.\,11805103 and No.\,11447204) and Jiangsu natural science foundation (Grant No.\,BK20190137).

\bibliographystyle{prsty}
\bibliography{article}
\end{document}